\documentclass{IEEEtran4PSCC}
\usepackage{multirow}
\ifCLASSINFOpdf
   \usepackage[pdftex]{graphicx}
\else
   \usepackage[dvips]{graphicx}
\fi
\usepackage[cmex10]{amsmath}
\usepackage{mathtools}
\usepackage{amssymb}

\usepackage{algorithmic}

\usepackage{array}
\usepackage{url}

\usepackage{soul}
\usepackage{flushend}

\makeatletter
\let\old@ps@headings\ps@headings
\let\old@ps@IEEEtitlepagestyle\ps@IEEEtitlepagestyle
\def\psccfooter#1{%
    \def\ps@headings{%
        \old@ps@headings%
        \def\@oddfoot{\strut\hfill#1\hfill\strut}%
        \def\@evenfoot{\strut\hfill#1\hfill\strut}%
    }%
    \def\ps@IEEEtitlepagestyle{%
        \old@ps@IEEEtitlepagestyle%
        \def\@oddfoot{\strut\hfill#1\hfill\strut}%
        \def\@evenfoot{\strut\hfill#1\hfill\strut}%
    }%
    \ps@headings%
}
\makeatother

\psccfooter{%
        \parbox{\textwidth}{\hrulefill \\ \small{21st Power Systems Computation Conference} \hfill \begin{minipage}{0.2\textwidth}\centering \vspace*{4pt} \includegraphics[scale=0.06]{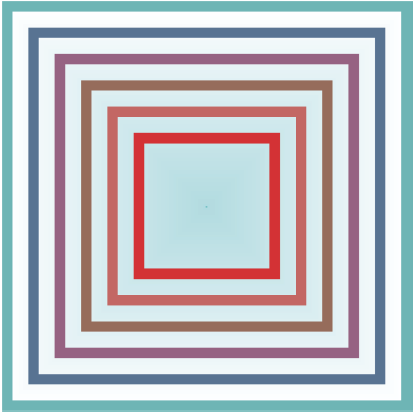}\\\small{PSCC 2020} \end{minipage} \hfill \small{Porto, Portugal --- June 29 -- July 3, 2020}}%
}

\begin{document}

\title{Bayesian Hierarchical Methods for Modeling Electrical Grid Component Failures}
%

\author{
\IEEEauthorblockN{Laurel N. Dunn, Ioanna Kavvada, Mathilde D. Badoual, and Scott J. Moura}
\IEEEauthorblockA{Department of Civil \& Environmental Engineering \\
University of California, Berkeley\\
Berkeley, CA\\
\{lndunn@berkeley.edu\}
}}


\maketitle

\begin{abstract}
Failure probabilities for grid components are often estimated using parametric models which can capitalize on operational grid data.
This work formulates a Bayesian hierarchical framework designed to integrate data and domain expertise to understand the failure properties of a regional power system, where variability in the expected performance of individual components gives rise to failure processes that are heterogeneous and uncertain.
We use Bayesian methods to fit failure models to failure data generated in simulation.
We test our algorithm by evaluating differences between the data-generating model, our Bayesian hierarchical model, and maximum likelihood parameter estimates.
We evaluate how well each model can approximate the failure properties of individual components, and of the system overall.
Finally, we define an upgrade policy for achieving targeted reductions in risk exposure, and compare the magnitude of upgrades recommended by each model.
\end{abstract}


\thanksto{This research was supported by a 2019 Seed Fund Award from CITRIS and the Bantao Institute at the University of California.}

\section{Introduction and Motivation}
Millions of individual components make up electric power systems; each of these has a finite operating life time.
Grid components are subjected to physical force, chemical processes, and operational stress that can give rise to failure.
Grid hardening and preventative maintenance can mitigate the risk of unexpected failures, but risk reduction measures can negatively impact cost and even performance.
High-fidelity models for representing failure processes in electric power systems can provide insights valuable to a number of resource allocation decisions, including system upgrades \cite{Nagarajan2016}, component stockpiling \cite{Coffrin2011}, and disaster response \cite{nateghi2018}.

The current work presents a framework for probabilistically modeling the risk that components will fail when exposed to stress.
This work provides a basis for leveraging grid data to detect and quantify unmitigated vulnerabilities in operational power systems, and to inform upgrade decisions.
We explore challenges that arise when using machine learning methods to characterize failure properties of real-world systems.
These challenges include heterogeneous performance characteristics of individual components (e.g., due to differences in state-of-health), and sparse observational data of past failure events.

The contributions of this work are twofold:

First, we propose a Bayesian hierarchical framework for capitalizing on data and domain expertise to model component-level and system-wide failure properties.
Our formulation captures the stochastic and heterogeneous nature of failure processes by assuming model parameters are uncertain and random.
Accounting for this randomness provides insight into the range of outcomes that could occur in light of the uncertainty that exists.
This feature makes our model particularly well-suited to examining the possibility of low-probability, high-impact risk scenarios that may not be evidenced in past data.
Applications could include scenario analysis of evolving risks due to climate change, including wildfires and severe weather events.

Second, we formulate a mathematical model for optimizing component upgrades to achieve specified risk thresholds.
This policy capitalizes on uncertainty in the failure parameters of the system for insight into the performance characteristics of existing components.
We describe how decision-makers can apply these insights to identify the subset of components with the highest probability of failure, thus minimizing the number of upgrades needed to achieve risk reduction goals.

This paper is structured as follows.
In Section \ref{sec:lit_review} we contextualize our work within the existing literature.
Section \ref{sec:formulation} motivates the use of Bayesian hierarchical models in power systems, and describes our model formulation.
Section \ref{sec:methods} describes methods for parameter estimation, model selection, and model evaluation--and formulates the optimal upgrade policy.
Section \ref{sec:results} compares the parameters, statistical properties, and decision-making implications of our model with alternative models for systems exhibiting a range of failure properties. 
Section \ref{sec:conclusions} presents concluding remarks and outlines opportunities for future work.

\section{Literature Review}
\label{sec:lit_review}

A number of studies in the literature examines failure properties of regional power systems during extreme events.
Some of these studies use prescriptive failure models to inform probabilistic models for assessing risk and weighing upgrade policies in simulation \cite{Nagarajan2016, panteli2017, ren2008}.
Other studies mine failure data from past events to characterize vulnerabilities in operational power networks, for example to cascading failures \cite{clarfeld2018}, hurricanes \cite{reed2016} and earthquakes \cite{park2006}.

A separate body of literature focuses on estimating the probability that individual components will fail when exposed to certain ambient or operating conditions.
It is conventional to use parametric failure models (or ``fragility curves'') to describe these failure probabilities; \cite{lallemant2015} provides a survey of models and methods that are typically used.
In power systems, standard failure models are available \cite{ieeec57}, and a number of studies report models characterizing failures that occurred under severe conditions such as hurricanes \cite{reed2016, Murray2014} and earthquakes \cite{park2006}.

The literature also suggests, however, that failure properties are not the same for all grid components.
For example, \cite{mohammadi2019} and \cite{rudin2010} mine existing data to characterize differences in failure probabilities based on what is known about components that failed in the past. 
However, grid data are often noisy and sparse, and the degradation mechanisms that give rise to failure may be costly (if not impossible) to monitor \cite{ieeec57}.

These studies underscore two critical research needs.
First, there is a need to better understand failure properties of grid components in light of the fact that grid data can be sparse, noisy, costly, and inaccurate (see \cite{rudin2010}).
Second, there is a need to examine if or how uncertainty in component-level failure models could change how we understand the risk of low-probability high-impact events.

\section{Model Formulation}
\label{sec:formulation}
Here, we formulate a Bayesian hierarchical model to characterize failure properties of grid components in a regional power system.
We begin by providing philosophical and mathematical context on Bayesian hierarchical models, and discuss why they are uniquely well-suited to the application at hand.
We go on to describe the functional form of the models we use to characterize failure probabilities for individual grid components, and draw a mathematical link between component-level failure probabilities and system-wide failure of regional power systems under stress.

\subsection{Bayesian Hierarchical Model}
\label{sec:bhm}
Hierarchical models provide a basis for characterizing systems where the relationship between the input and output variables is probabilistic and uncertain.
The model is structured as a multi-level hierarchy where the last level describes some probabilistic process whose outcome is conditioned on the outcomes at previous levels in the hierarchy.
Thus levels of the model describe uncertainty in the outcome, in the data, in the system of equations relating the inputs to the outputs.

To formalize this mathematically, let us consider a system where the output of the system $y$ is related to some input $x$ by a given model $M_\theta(x)$ with parameters $\theta$, such that
\begin{align}
    y = M_\theta(x)
\end{align}
If the input $x$ is a realization of some random variable $X$, then the output $Y$ is also a random variable which is conditionally dependent on $X$.
The probability of observing a particular set of outcomes $P(x,y)$ can be written
\begin{align}
\label{eq:hm}
P(x,y) = P(y|\theta,x)\,\,P(x)
\end{align}
\noindent where the outcome $y$ depends on the realization of $x$ that was observed, and on the parameters $\theta$ given.
This formulation can be extended to account for additional sources of uncertainty, for example in how the system is parameterized.

Hierarchical models capitalize on this structure of conditional probabilities to characterize the likelihood that a particular parameterization is correct, in light of the observations $x$ and $y$ available to us.
Inherent to this approach is the notion that parameter estimates can only be as definitive as the data that are available to compute them.
When the data are sparse, or when the mapping of $X$ onto $Y$ is imprecise (e.g., due to measurement noise, or unobservable system dynamics), then parameter estimates will be uncertain.

To demonstrate this mathematically, we return to \eqref{eq:hm} above.
Suppose we wish to estimate the probability $P(\theta|x,y)$, or the distribution of unknown parameters $\theta$, given an observation of $x$ and $y$.
The Bayesian formulation is as follows:
\begin{align}
    P(\theta|x,y) = \frac{P(\theta)\,\,P(x,y|\theta)}{P(x,y)}
\end{align}
Here, $P(\theta)$ is the prior, which allows us to incorporate expert judgement into our calculation $P(\theta|x,y)$ (i.e., the posterior) in addition to the data $x$ and $y$.
Here, the term $P(x,y|\theta)$ is the ``likelihood function'' describing the probability of observing the data $(x,y)$ given a particular estimate of the parameters. Finally, the denominator $P(x,y)$ is the likelihood of observing the data independent of the parameters.
Using the law of total probability, we can rewrite the denominator as
\begin{align}
\label{ref:normalizing_constant}
P(x,y) = \int_{\theta' \in \Theta} P(x,y|\theta') P(\theta') d\theta'
\end{align}
This quantity is the ``normalizing constant'' and generally cannot be computed analytically.
Instead, we use Markov Chain Monte Carlo methods (described in Section \ref{sec:mcmc}) to compute the posterior numerically.

\subsection{Component Fragility Curves}
\label{sec:fragility}
Fragility curves are commonly used in reliability engineering to describe the probability that a system will fail when exposed to different magnitudes of stress.
These curves provide a quantitative basis for evaluating how vulnerable a system is to different modes of stress, and to stress conditions that are rarely (if ever) observed in practice.
We refer readers to \cite{lallemant2015} for a detailed overview of fragility analysis.

We use a generalized linear model with a logistic link function relating the stress input conditions $X$ to failure occurrence.
We define $g(x)$ to be the probability that a particular component will fail, given stress conditions $x$, written mathematically as
\begin{eqnarray}
    g(x) &=& P(\text{``component fails''}|x) \nonumber \\
         &=& \frac{1}{1+\exp \left[-\sum_{i=1}^p \beta_i (x_i-\alpha_i)\right]} \label{eq:logit}
\end{eqnarray}
Here $x_i$ is a vector of time series observations for a particular stress condition.
Elsewhere in the text, we refer to $X$---a p-dimensional matrix of measurement data.
The thresholds $\alpha_i$ describe the ability of the system to withstand mild to moderate stress conditions.
The coefficients $\beta_i$ describe how quickly the failure probability increases as stress conditions approach (and exceed) the threshold $\alpha_i$.
These thresholds and slopes parameterize the failure mechanics of the system, providing a probabilistic relationship between the stress conditions $X$ and failures.
We refer to the parameters collectively as $\theta$.


\subsection{System-Wide Failure Properties}
\label{sec:failures}
Here, we extend the component-level failure model in \eqref{eq:logit} to model system-wide failure.
Power systems are designed and operated to be resilient to the loss of one or two components.
However, losing additional components can cause the system to become inoperable, even when only a small fraction of components are affected.
Since the ambient conditions that give rise to failures (e.g., storms, heat waves, etc.) typically occur over a time-span of hours to days, the probability that any one specific component will be lost to the system is small (component lifespans are typically on the order of decades).

At the component level, failure occurrence is a binary random variable, and can be modeled as a Bernoulli process with failure probability $g(x)$ (see Section \ref{sec:fragility}).
It follows that from a systems perspective, the overall number of failures $Y$ can be modeled as a sequence of Bernoulli trials.
The number of trials is simply equal to the number of components in the system $N$, and $Y$ follows a Binomial distribution where $Y\sim Binomial(N\times g(x))$.

The Poisson limit theorem tells us that if $N$ is large, a Binomial random variable can be approximated by a Poisson process with rate $\lambda = N\times g(x)$.
The probably of observing a particular number of failures $y$ is
\begin{align}
\label{eq:poisson}
    P(y) = e^{-N\times g(x)} \frac{(N\times g(x))^y}{y!}
\end{align}
To put this in words: the system-wide failure process $Y$ can be approximated by a Poisson process, where the rate is related to the failure probabilities of the components in the system as specified in \eqref{eq:logit}.

Having formalized the failure probability distributions, we revisit our description of the model $M_{\theta}(x)$.
We define $M_\theta$ to be a function mapping ambient conditions $x \in \mathbb{R}^{p}$ onto a random variable $Y$ describing the number of failures in the system.
The variable $Y$ is an inhomogenous Poisson process; the observed failures $y$ are a realization of $Y$.
Mathematically, $M_\theta(x)$ is defined as:
\begin{align}
\label{eq:link}
    M_\theta&: x \mapsto Y \\
    Y &\sim \text{Poisson}\left(\frac{N}{1+\exp\left[\sum_{i=1}^p -\beta_i (x_i - \alpha_i)\right]}\right)
\end{align}

\section{Analysis Methods}
\label{sec:methods}
\begin{figure}
    \centering
    \includegraphics[width=0.45\textwidth]{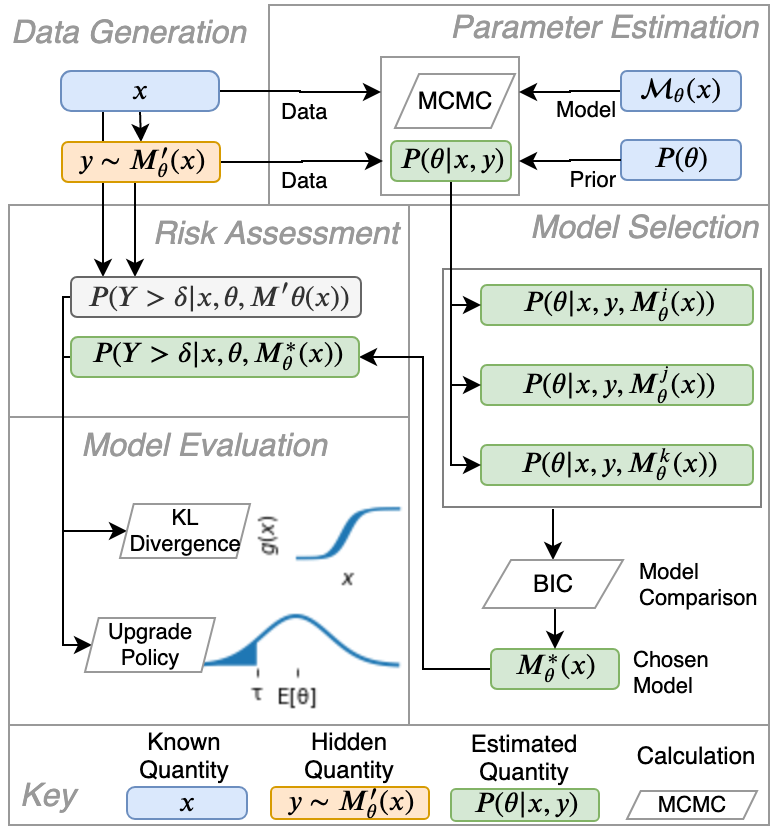}
    \caption{Block diagram illustrating analysis procedure.}
    \label{fig:flowchart}
\end{figure}

Fig. \ref{fig:flowchart} provides a visual overview of the framework we use to fit and evaluate our Bayesian hierarchical model.
The following paragraphs provide details about methods for estimating parameters given a candidate model $M^{i}_\theta(x)$, for choosing the best model $M^*_\theta(x)$ from a library of options $\mathcal{M}_\theta(x)$, and for benchmarking the performance of our model against others.

\subsection{Parameter Estimation}
\label{sec:mcmc}
We use Markov Chain Monte Carlo (MCMC) methods to compute the posterior distribution of the parameters $\theta$.

MCMC encompasses a category of algorithms that approximate an unknown probability distribution by sampling candidate values from some proposed distribution, and then updates the proposed distribution at each iteration.
Updates are made according to some rule based on the likelihood function.
In our case, we examine the likelihood of the data given a candidate set of parameters $\theta^*$, given by $P(x,y|\theta^*)$.

Different MCMC algorithms use different rules for updating the sampling distribution of $\theta^*$ at each iteration.
However, these rules are all designed such that the sampling distribution asymptotically converges to a target distribution.
Convergence is achieved either by updating the sampling distribution at each iteration, or using rejection sampling.
We use the common Metropolis-Hastings algorithm \cite{metropolis1953} which relies on rejection sampling.
We refer readers to \cite{smith1993} and \cite{roberts2001} for details on implementation.

At each iteration $k$, updated parameters $\theta^{*}$ are proposed according to the following rule:
\begin{equation}
    \theta^* = \theta^k + z
\end{equation}
where $z$ is a random perturbation sampled from a multivariate normal distribution.

As the chain advances, the parameters $\theta^{k+1}$ are set equal to $\theta^*$ if certain acceptance criteria are satisfied.
If the criteria are not met, the new parameters are rejected and the current parameters are kept (i.e., $\theta^{k+1}$ is set equal to $\theta^{k}$).
The acceptance rule is given as follows:
\begin{equation}
\label{eq:metropolis_hastings}
P(\text{accept}) = \min \left\{1, \frac{P(\theta^{*}|y)}{P(\theta^{k}|y)}\right\}
\end{equation}
In words: if the likelihood of the new parameters $P(\theta^*|y)$ exceeds the likelihood of the current parameters $P(\theta^{k}|y)$, the update is accepted.
If the proposal decreases the likelihood function, then the update is accepted with some probability equal to the ratio between the two likelihoods.

We use empirical methods to generate the prior and to determine the correlation structure of $Z$: the so-called ``jumping distribution''.
We do so by computing maximum likelihood parameter estimates $\hat{\theta}$ for  bootstrapped sub-samples of the data.
We then adjust the covariance of $Z$ by some scalar value to achieve an acceptance rate of 0.25 to ensure that the Markov chain exhibits desirable convergence properties \cite{roberts2001}.

\subsection{Model Selection}
We use the Bayesian Information Criterion (BIC) to weigh different parametric models $\mathcal{M}_\theta(x)$ based on parsimony and fit.
Mathematically, the BIC is given by:
\begin{align}
    \label{eq:BIC}
    BIC_{M^i_{\theta}(x)} = -2 \log \left(\ell_{M^i_\theta}\right) + K \log(n)
\end{align}
Here, $\ell_{M^i_\theta}$ describes the likelihood that a particular model $M^i_\theta(x)$ describes the data, $K$ is the number of features in the model, and $n$ is the number of observations in the data.
The likelihood is calculated by taking the total probability of the model across all candidate parameterizations $\theta'$, given by
\begin{align}
\ell_{M_\theta} = \int_{\theta' \in \Theta} P(x,y|M_\theta, \theta') P(\theta') d\theta'    
\end{align}

The BIC describes the information contents of one model compared with another.
Thus, we choose the model that minimizes the BIC, and compare the information contents of different models based on the differences between the BIC for one model compared with the model selected.
We refer readers to \cite{posada2004} for a comparison of BIC and other metrics for model selection.

\subsection{Evaluating Model Performance}
Here, we describe metrics for evaluating how accurately the model learned from data recovers the true model.

In real-world systems, the true model would be unknown.
In simulation, however, we have specified the true model in order to generate the data $y$.
We capitalize on the fact that the failure model is known to evaluate the performance of our Bayesian hierarchical model.
We evaluate the performance of our model by examining how accurately it represents the statistical properties of the system (measured using the K-L divergence), and in terms of its ability to recommend suitable upgrade policies.

\subsubsection{Kullback-Leibler (KL) Divergence}
KL divergence is an information theoretic metric describing the statistical distance between two distributions, given by:
\begin{align}
\label{eq:divergence}
    D_{KL}(P||Q) = \int_{y \in \mathcal{Y}} P(y) \left[log\left(P(y)\right) - log\left( Q(y)\right)\right] dy
\end{align}
Here $P(y)$ is the probability of observing $y$ failures given the true distribution of $\theta$
\begin{align}
\label{eq:divergence_integral}
    P(y) = \int_{x \in \mathcal{X}} \int_{\theta \in \Theta} P(y|x,\theta) P(x) P(\theta) \, d\theta\, dx
\end{align}
while $Q(y)$ is the probability of observing $y$ failures given the distribution of $\hat{\theta}$ approximated using MCMC.

\subsubsection{Optimal Upgrade Policy}
\label{sec:upgrades}
Next, we examine upgrade policies that a decision-maker would arrive at using the proposed model, compared with the true model. 
The objective of the upgrade policy is to identify the minimum number of components that must be be upgraded to achieve a target distribution of acceptable failure rates.

We define the target distribution $P(\Tilde{Y})$ such that the probability that $\tilde{Y}$ failures exceeds some specified damage threshold $\delta$ must be below a specified probability $\epsilon$:
\begin{align}
\label{eq:upgrade_policy}
    P\left(\Tilde{Y}>\delta\right) &\le \epsilon \\
    P\left(\Tilde{Y}>\delta\right) &= 1-\sum_{\tilde{y}=0}^\delta P(\tilde{Y} = \tilde{y})
\end{align}
where $P(\tilde{Y} = \tilde{y})$ is given by \eqref{eq:poisson}.
For example, the policy could be to upgrade the system such that damages exceeding $>1\%$ of components are expected to happen no more than once every 100 years.

The optimal policy finds the minimum number of components $M$ that must be upgraded to achieve the target distribution $P(\Tilde{Y})$.
Upgrade decisions modify the parameters $\theta_i$ of individual components, thus changing the distribution of parameters in the overall system $P(\Tilde{\theta})$.

In Section \ref{sec:formulation}, we introduce the notion that uncertainty in the parameters of the system stems from variability in the parameters of individual components.
The implication is that the distribution $P(\theta)$ provides meaningful insight into the failure probabilities of individual components, and that the optimal upgrade policy would target the subset of components that are at highest risk of failure (given $x$).
Though the failure parameters of individual components may not be known, exogenous information about the age or state-of-health of individual components are often available.
We assume that decision makers can capitalize on these data to accurately assess the relative performance of one component verses another.

\begin{figure}
    \centering
    \includegraphics[width=0.45\textwidth]{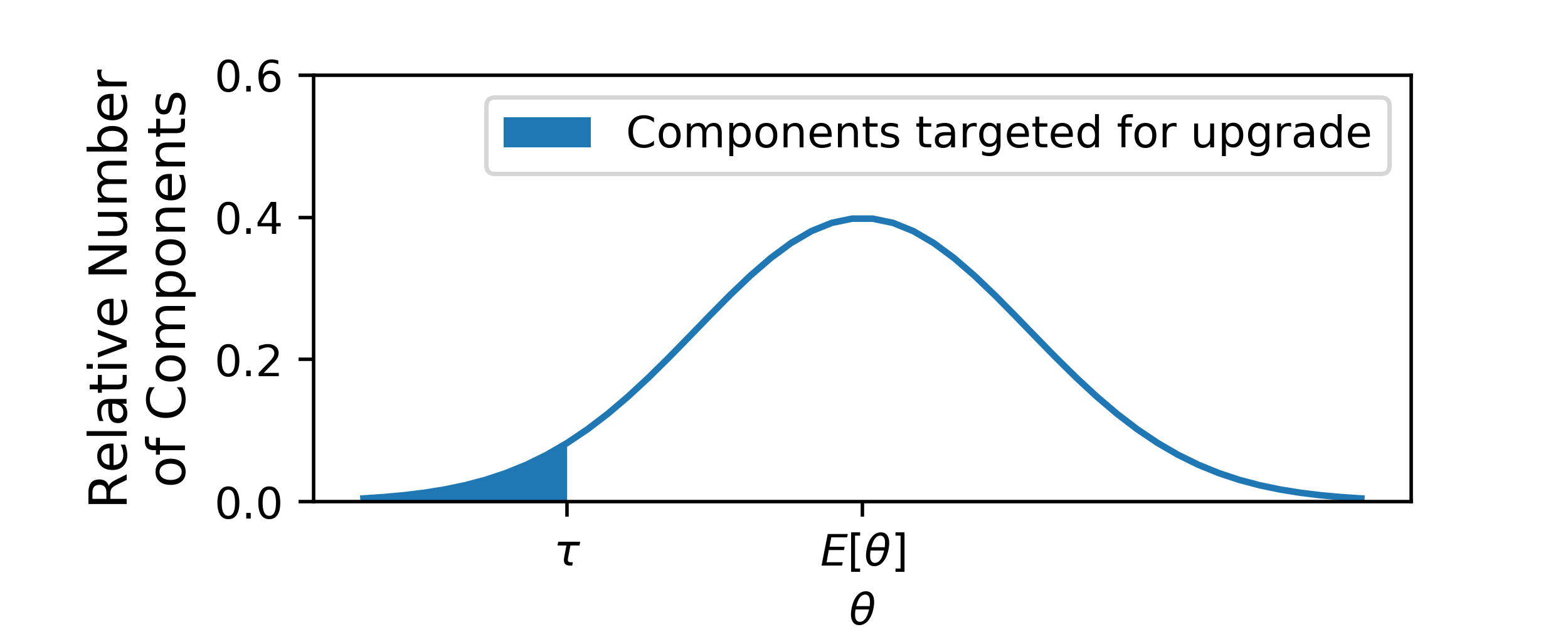}
    \caption{Illustrative diagram showing failure parameters for components targeted for upgrades. The number of components is given by \eqref{eq:n_upgrades}. The threshold $\tau$ is obtained by solving \eqref{eq:upgrade_policy}}.
    \label{fig:upgrades}
\end{figure}

We consider an upgrade policy where upgrades increase the threshold parameters $\alpha$, focusing on a particular case study where upgrade decisions would increase the wind speed rating of poles.
The optimal upgrade policy for such a measure is depicted graphically in Fig. \ref{fig:upgrades}.
Thus the target distribution $P(\Tilde{\theta})$ is simply a truncation of the original distribution $P(\theta)$, where upgrades are targeted to replace components with relatively lower failure thresholds.

The optimization objective is to identify the truncation threshold $\tau$.
Mathematically, we can compute $\tau$ by numerically solving the following:
\begin{align}
\label{eq:n_upgrades}
    \epsilon \ge 1- \sum_{y=0}^\delta \int_{x \in \mathcal{X}} \int_{\theta = \tau}^\infty P(\tilde{y}|x,\theta)\,P(x)\,P(\theta)\,d\theta\,dx
\end{align}
As presented here, this formulation assumes that the failure probability of upgraded components is zero.

We assume that the parameters of upgraded components are modified such that the failure probability becomes negligible.
Once $\tau$ is known, the optimal number of components to upgrade can be computed from the integral:
\begin{align}
    M = N \int_0^\tau P(\theta)\,d\theta
\end{align}

\section{Results and Discussion}
\label{sec:results}

Here, we present results from a series of tests designed to evaluate how accurately the Bayesian Hierarchical model (BHM) detailed in Section \ref{sec:formulation} and \ref{sec:methods} approximates failure processes representative of real power systems.

We train a model using synthetic failure data generated in simulation, where the parameters of the failure model are specified.
In the following paragraphs, we outline our procedure for generating the data, and describe the characteristics of the failure processes we examine.
We report results of the parameter estimation, model selection, and model evaluation procedures described in Section \ref{sec:methods}.
Where relevant, we compare results against the true model used to generate the data, and with a model trained using traditional maximum likelihood estimation (MLE), as described in \cite{lallemant2015}.

\subsection{Data Generation}
We generate synthetic failure data by defining a failure model that relates ambient conditions $x$ to the random variable $Y$.
Data are generated from a model where $x$ is a vector of wind speed measurements.
For $x$ we use 10-years of hourly weather data recorded at a site in Northern California to represent stress conditions on the system.

Given the weather conditions at each time step and the specified failure model $M'_\theta(x)$, we compute the failure probability $g(x)$ using \eqref{eq:link}.
From these component-level failure probabilities we generate a realization $y$ of the number of failures in the system overall.
The random variable $Y$ follows the Poisson distribution specified in \eqref{eq:poisson}.

\subsection{Experimental Design}
We test our method given failure processes that exhibit a range of statistical properties.

\begin{figure}
    \centering
    \includegraphics[width=0.5\textwidth]{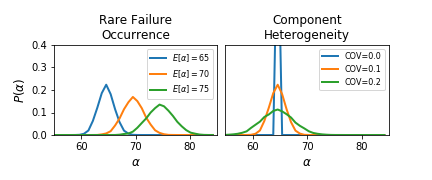}
    \caption{Probability density of threshold parameter $\alpha$ for different failure model specifications. Here, $\theta$ describes the wind speed (in m/s) at which the failure probability $g(x)$ is 0.5.}
    \label{fig:scenarios}
\end{figure}

First, we examine cases where components are increasingly robust to failure.
Because the probability $P(x)$ decreases as ambient conditions $x$ become increasingly severe, robust performance can be represented by increasing the parameter $\alpha$ in \eqref{eq:logit}.
Failure models are challenging to learn for components that are robust to failure because as $P(x>\alpha)$ decreases, the probability of observing failure events becomes small.
We examine cases where $P(\alpha)$ is normally distributed and is centered around three wind speed thresholds (65, 70, and 75 m/s), as depicted in Fig. \ref{fig:scenarios}.

Next, we examine cases where failure properties of different components in the system exhibit varying degrees of heterogeneity.
We represent heterogeneity by increasing the coefficient of variation (COV) in failure parameters, or the ratio between the standard deviation and the mean.
A higher COV indicates more variability in the performance characteristics of individual components.

We execute parameter estimation, model selection, and model evaluation using the methods discussed in Section \ref{sec:methods} for each of the models specified.
We evaluate how well our model compares with the true model.
We also compare against a model of the same functional form as the selected model $M^*_{\theta}(x)$ that uses maximum likelihood estimation to estimate the parameters of the model (rather than using MCMC to estimate the posterior). 
That is, $M^*_{\theta}(x)$ estimates scalar values for the parameters as opposed to probability distributions.

\subsection{Results and Discussion}
Here, we summarize results for each of the failure conditions specified above.
We use the BIC to select the functional form of $M^*_\theta(x)$ from candidate models where failures are correlated with only wind speed, or both wind speed and precipitation.
In all of the cases we consider, we find that minimizing the BIC results in selecting the correct model.

Next, we focus our analysis on examining estimates of the threshold parameter $\alpha$ in particular, where $\alpha$ describes the wind speed $x$ at which $g(x)$ equals 0.5.
The following paragraphs examine the magnitude and distribution of the parameter estimates fitted to failure data generated for each of the specified failure models. 
We also report the K-L divergence and upgrade policy, as defined in Section \ref{sec:methods}.

\subsubsection{Parameter Estimates}
\begin{figure}
    \centering
    \includegraphics[width=0.5\textwidth]{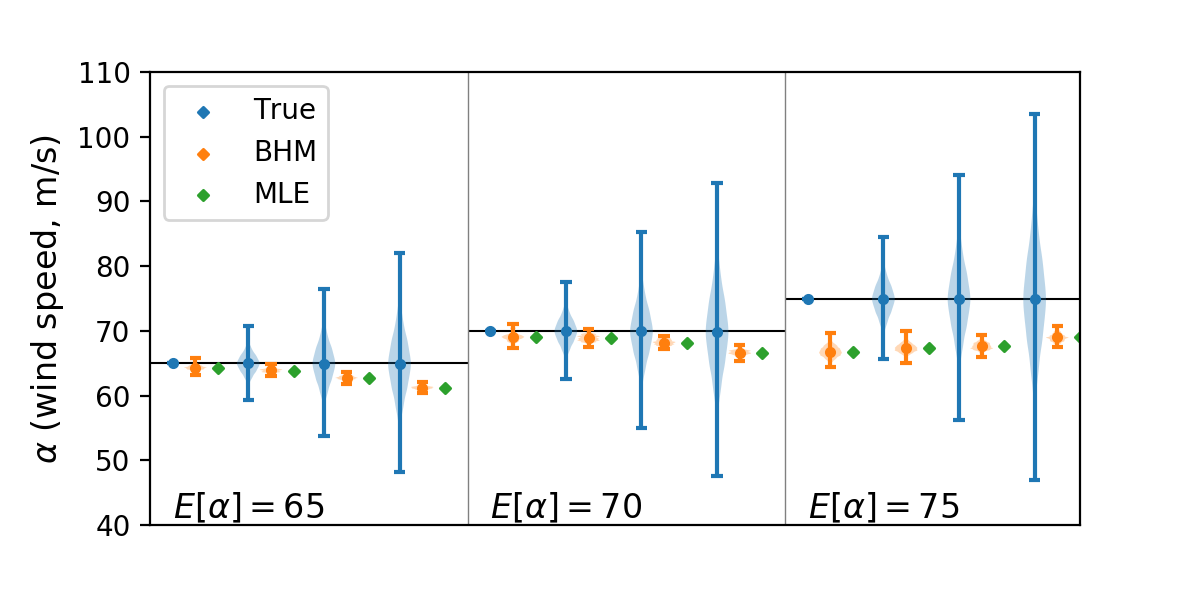}
    \caption{Violin plot showing range and distribution of true values, and BHM and MLE estimates of $\alpha$. Horizontal lines denote the mean of the true distribution $E\left[\alpha\right]$. From left to right, the COV for parameters in each panel are 0, 0.1, 0.2, and 0.3.}
    \label{fig:violins}
\end{figure}

Fig. \ref{fig:violins} shows the range and distribution of $\alpha$ for the true model, and for the BHM and MLE approximations.
Both the BHM and MLE models produce mean values within $<5\%$ of the true value when the true mean $E\left[\alpha\right]$ is 65 and 70 m/s, and within $<10\%$ when the true mean is 75 m/s.
However, in all cases the parameter distributions are systematically lower than the true mean.
This difference increases in cases where the variance is high.

This result suggests that parameter estimates are systematically baised.
The reason for this bias is that the probability of observing damaging wind speed conditions is higher when the threshold $\alpha$ is low.
It follows that the incidence of failures is relatively higher among components with small $\alpha$ than among components with high $\alpha$.
The implication is that parameter estimates are more heavily influenced by a relatively small subset of components that particularly prone to failure.

The estimated coefficients $\beta$ for each case are listed in Table \ref{tab:betas}.
In our model, the coefficients are taken to be constant and are equal to 0.2.
BHM estimates report very narrow uncertainty bounds on slope parameter estimates.
We note that in cases when the wind speed threshold is high, $\beta$ tends to decrease as the COV increases.
This result has implications on upgrade decisions, as discussed in detail below.

\begin{table}
\renewcommand{\arraystretch}{1}
    \centering
    \begin{tabular}{c|c|c|c|c}
     & COV=0 & COV=0.1 & COV=0.2 & COV=0.3 \\ \hline
    $E\left[\alpha\right]=65$ & 0.201 & 0.201 & 0.203 & 0.206 \\
    $E\left[\alpha\right]=70$ & 0.201 & 0.201 & 0.199 & 0.197 \\
    $E\left[\alpha\right]=75$ & 0.239 & 0.232 & 0.220 & 0.199 \\
    \end{tabular}
    \caption{Estimated wind speed coefficient $\beta$ for BHM and MLE models. In the true model, $\beta$ is always 0.2.}
    \label{tab:betas}
\end{table}

\subsubsection{K-L Divergence}
\begin{figure}
    \centering
    \includegraphics[width=0.4\textwidth]{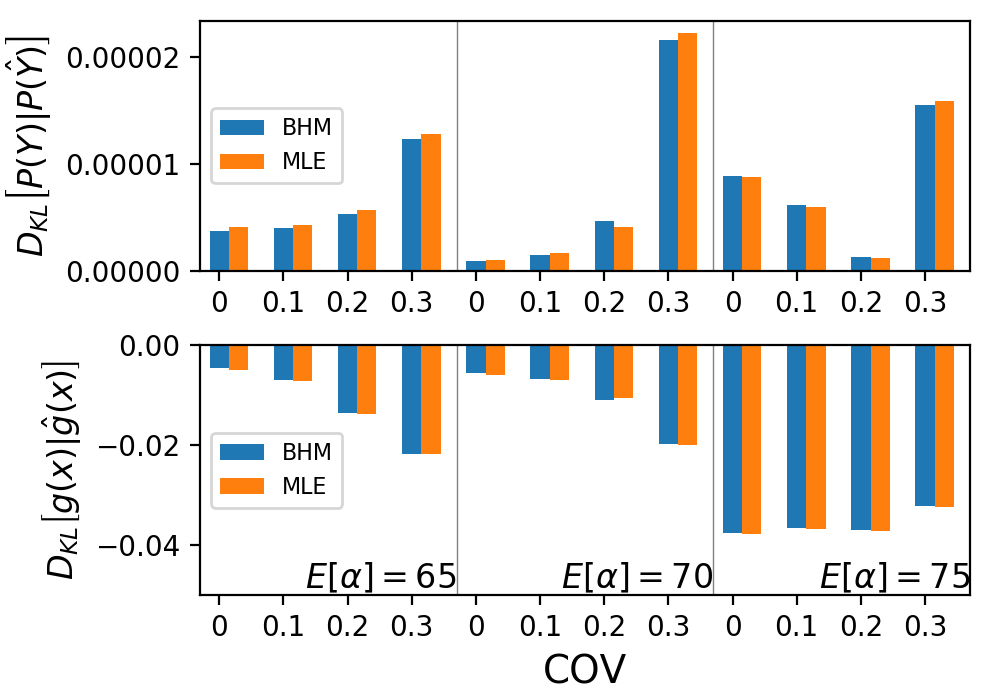}
    \caption{Statistical distance between the true distribution $P(Y)$ and estimated distribution $P(\hat{Y})$ of failures to the system (top), and between the true failure probability $g(x)$ and estimated failure probability $\hat{g}(x)$ of individual components.}
    \label{fig:kl_divergence}
\end{figure}

Fig. \ref{fig:kl_divergence} shows the statistical distance between the true and estimated probabilities $P(Y)$ (top) and $g(x)$ (bottom), as given by \eqref{eq:divergence}.
The true distribution is omitted, as the distance is by definition zero.
The K-L divergence is computed by taking the integral across all parameter values $\theta \in \Theta$ (as indicated in \eqref{eq:divergence_integral}), and can be thought of as a metric of error in the underlying probability distributions.

Results show that there is little difference between the BHM and MLE models compared with the true model.
When the mean threshold is 65 or 70 m/s, the divergence increases as heterogeneity among parameters increases.
This result does not hold when the threshold is 75 m/s, though it is unclear to what extent this result is driven by the particular realization of failures $y$ that occurred.

At the component level, the divergence is consistently negative--indicating that both the BHM and MLE models under-represent component-level failure probabilities.
This result is consistent with the previous observation that both models under-estimate the threshold parameter $\alpha$.
When the failure threshold is high, the statistical error in component-level failure models is high, though error in the distribution of failures in the system overall may be low.
Future work will explore how sensitive these results are to the particular realization of failures $y$ that occurred.

\subsubsection{Upgrade Policy}
Finally, we compare differences in upgrade decisions informed by the true model, and by BHM and MLE parameter estimates.
We use the upgrade model detailed in Section \ref{sec:upgrades}, where the risk tolerance is defined as follows:
\begin{align}
    P(Y>0.1N) \le 0.05
\end{align}
In words, our objective is to limit the probability that $>10\%$ of components will be damaged (i.e., $\delta=0.1N$); we aim to limit the probability to below 5\% (i.e., $\epsilon=0.05$).

From \eqref{eq:n_upgrades}, it is clear that upgrades are related not only to the failure model, but also to the ambient conditions $x$.
Here, we take $Y$ to be the total number of failures in a year, and take $P(x)$ to be a log-normal distribution fitted to 10 years of historic data.

\begin{figure}
    \centering
    \includegraphics[width=0.5\textwidth]{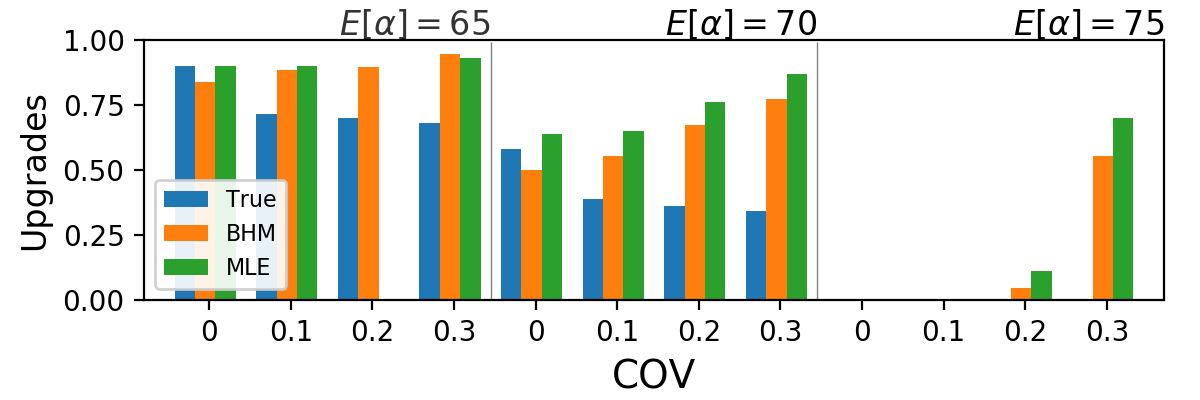}
    \caption{Percentage of components recommended for upgrade.}
    \label{fig:upgrade_results}
\end{figure}

Fig. \ref{fig:upgrade_results} summarizes the number of upgrades necessary to achieve this risk target for each of the cases we consider.
We also show the threshold $\tau$ that defines the subset of components targeted for upgrades.
Systems where $P(\alpha)$ is centered around 65 or 70 m/s do not meet the specified risk target given the wind speed conditions that are present, and the model recommends substantial upgrades to achieve the specified risk threshold.
On the other hand, when $P(\alpha)$ is centered around 75 m/s, the system is overbuilt--and no components are targeted for upgrades.

In general, the BHM formulation outperforms the MLE model in recommending upgrades that are less aggressive and are more closely aligned with the true model.
The reason is that by representing variability in the failure parameters of the components, the BHM can more accurately represent the fact that the benefits to the system are higher when upgrade policies target components with the highest failure probability.

Errors in upgrade policies informed by the BHM and MLE models tend to favor more aggressive upgrades than are necessary.
This result is consistent with our previous result that these models systematically underestimate the mean threshold $\alpha$.
What is less intuitive is the observation that both models recommend aggressive upgrade policies in the case where $E\left[\alpha\right]$ is 75 m/s and the variance is high.
Returning to Table \ref{tab:betas}, we can see that estimates of $\beta$ are lower in these systems.
One consequence of underestimating $\beta$ is that the failure probability $g(x)$ increases for $x<\alpha$.
In other words, underestimating $\beta$ causes us to overstate the risk of failures under mild ambient conditions, resulting in an unnecessarily aggressive upgrade policy.


\section{Concluding Remarks}
\label{sec:conclusions}
This paper outlines a Bayesian hierarchical modeling framework designed to capitalize on grid data of historic failures to characterize failure probabilities and inform upgrade decisions.
We train our model on synthetic failure data generated from specified failure models, and benchmark our results against both the true model and against a model fitted using the maximum likelihood estimation.

When we assume that components exhibit heterogeneous failure properties, we find that our model tends to systematically under-represent variability relative to the true model.
Furthermore, we find that bias in the failures recorded in the data leads to systematic bias in the parameters of both Bayesian and maximum likelihood approximations.

This finding suggests a critical flaw in current best-practices for evaluating risk in electric power systems, and motivates the need for future work.
One opportunity to advance the work would explore sampling techniques--such as stratified MCMC sampling or importance sampling--which can adjust for known bias in the sampling distribution of the data (as the Metropolis Hastings algorithm does not).
Another opportunity is to explore the value of information that could help to more precisely determine the failure properties of components that do (or do not) pose a risk, in light of the sampling bias that exists.

\bibliographystyle{IEEEtran}
\bibliography{refs}

\end{document}